%% file: main.tex
\documentclass[conference]{IEEEtran}
\IEEEoverridecommandlockouts
\usepackage{fancyhdr}
\usepackage{cite}
\usepackage{amsmath,amssymb,amsfonts}
\usepackage{algorithmic}
\usepackage{graphicx}
\usepackage{textcomp}
\usepackage{xcolor}
\usepackage{bm}
\usepackage{booktabs}
\usepackage[ruled,linesnumbered]{algorithm2e}
\usepackage{multirow}
\usepackage{enumitem}
\usepackage{microtype}
\usepackage{graphicx}
\usepackage{subfigure}
\usepackage{booktabs} 
\usepackage{amsfonts} 
\usepackage{amssymb}
\usepackage{amsmath}
\usepackage{sansmath}
\usepackage{MnSymbol,wasysym}
\usepackage{subfigure}
\usepackage{amsthm}
\usepackage{tcolorbox}
\usepackage{url}
\newtheorem{theorem}{Theorem}
\newtheorem{definition}[theorem]{Definition}

\newtheorem{proposition}[theorem]{Proposition}

\newcommand{\eat}[1]{}

\def\BibTeX{{\rm B\kern-.05em{\sc i\kern-.025em b}\kern-.08em
    T\kern-.1667em\lower.7ex\hbox{E}\kern-.125emX}}
\begin{document}

\clearpage

\title{Robust Attributed Graph Alignment via Joint Structure Learning and Optimal Transport}

\author{Jianheng Tang$^{2\dag}$, Weiqi Zhang$^{2}$, Jiajin Li$^{3}$, Kangfei Zhao$^{4}$, Fugee Tsung$^{1,2}$, Jia Li$^{1,2*}$\thanks{$^*$Corresponding author.} \thanks{$^{\dag}$Work done during an internship at Tencent AI Lab.}\\
$^{1}$Hong Kong University of Science and Technology (Guangzhou),\\
$^{2}$Hong Kong University of Science and Technology,\\
$^{3}$Stanford University,
$^{4}$Tencent AI Lab\\
\{jtangbf,wzhangcd\}@connect.ust.hk, jiajinli@stanford.edu, zkf1105@gmail.com, \{season,jialee\}@ust.hk 
}

\pagestyle{plain}    

\maketitle

\begin{abstract}

Graph alignment, which aims at identifying corresponding entities across multiple networks, has been widely applied in various domains. As the graphs to be aligned are usually constructed from different sources, the inconsistency issues of structures and features between two graphs are ubiquitous in real-world applications. Most existing methods follow the ``embed-then-cross-compare'' paradigm, which computes node embeddings in each graph and then processes node correspondences based on cross-graph embedding comparison. However, we find these methods are unstable and sub-optimal when structure or feature inconsistency appears. To this end, we propose SLOTAlign, an unsupervised graph alignment framework that jointly performs Structure Learning and Optimal Transport Alignment. We convert graph alignment to an optimal transport problem between two intra-graph matrices without the requirement of cross-graph comparison. We further incorporate multi-view structure learning to enhance graph representation power and reduce the effect of structure and feature inconsistency inherited across graphs. Moreover, an alternating scheme based algorithm has been developed to address the joint optimization problem in SLOTAlign, and the provable convergence result is also established. Finally, we conduct extensive experiments on six unsupervised graph alignment datasets and the DBP15K knowledge graph (KG) alignment benchmark dataset. The proposed SLOTAlign shows superior performance and strongest robustness over seven unsupervised graph alignment methods and five specialized KG alignment methods. \footnote{Code and data are released at \url{https://github.com/squareRoot3/SLOTAlign}}

\end{abstract}


\input{sec1_intro.tex}

\input{sec2_preliminary.tex}
\input{sec3_analysis.tex}
\input{sec4_method.tex}
\input{sec5_experiment.tex}

\input{sec6_relatedwork.tex}

\section{Conclusion}\label{sec7}
In this paper, we propose SLOTAlign, the first robust unsupervised method tailored to tackle the structure and feature inconsistency issues in graph alignment. Instead of following the embed-then-cross-compare paradigm, we approach the graph alignment task via intra-graph structure modeling and cross-graph optimal transport alignment in a unified manner. Then, we present a provably convergent alternating type algorithm to address the joint optimization problem. Extensive experiments demonstrate that SLOTAlign can outperform the state-of-the-art graph alignment and KG alignment methods by an up to 15\% absolute improvement in Hit@1 with a shorter running time, and is also the most robust model against structure and feature inconsistency.

\section*{Acknowledgements}
The research of Li was supported by NSFC Grant No. 62206067, Tencent AI Lab Rhino-Bird Focused Research Program RBFR2022008 and Guangzhou-HKUST(GZ) Joint Funding Scheme. The research of Tsung was supported by the Hong Kong RGC General Research Funds 16216119 and Foshan HKUST Projects FSUST20-FYTRI03B.

\bibliographystyle{plain}

\bibliography{myref}
\end{document}

%% file: sec1_intro.tex
\section{Introduction}\label{sec1}

Graph alignment refers to the problem of identifying the node correspondences (i.e., anchor links) across different graphs. With graph data becoming ubiquitous in the Web era, graph alignment establishes connections between multiple networks and integrates them into a world-view network for subsequent analysis and downstream applications. Thus, graph alignment provides a comprehensive perspective for structured data compared with mining each individual network. As a well-established problem, graph alignment has received much attention due to its vast applicable tasks, e.g., linking accounts in different social network platforms \cite{li2019partially, li2018distribution, mu2016user}, matching entities across different knowledge graphs \cite{sun2020benchmarking, zhao2020experimental, liu2022selfkg, zeng2022entity}, integrating protein-protein interactions of different species \cite{liu2017novel, kazemi2016proper}, merging scholar profiles of academic collaboration networks \cite{tang2008arnetminer, zhang2021balancing}.

Graph alignment is usually treated as a supervised problem \cite{zhang2015cosnet, liu2016aligning, man2016predict,zhang2021balancing}, in which a set of ground-truth node correspondences is given. However, these correspondences are usually unavailable and further suffer from the labor expensiveness issue in real-world applications. Thus, unsupervised graph alignment methods have attracted increasing attention \cite{heimann2018regal, gao2021unsupervised, liang2021unsupervised, chen2019unsupervised, derr2021deep, zhou2021unsupervised, karakasis2021joint}. Also, graph nodes are often associated with wealthy side information, such as the user information of social network accounts or the embedding of knowledge graph entities. These high-dimensional node features/attributes can serve as an additional source of knowledge in graph alignment, especially under the unsupervised setting.

\begin{figure}[t!]
	\centering
	\includegraphics[width=\linewidth]{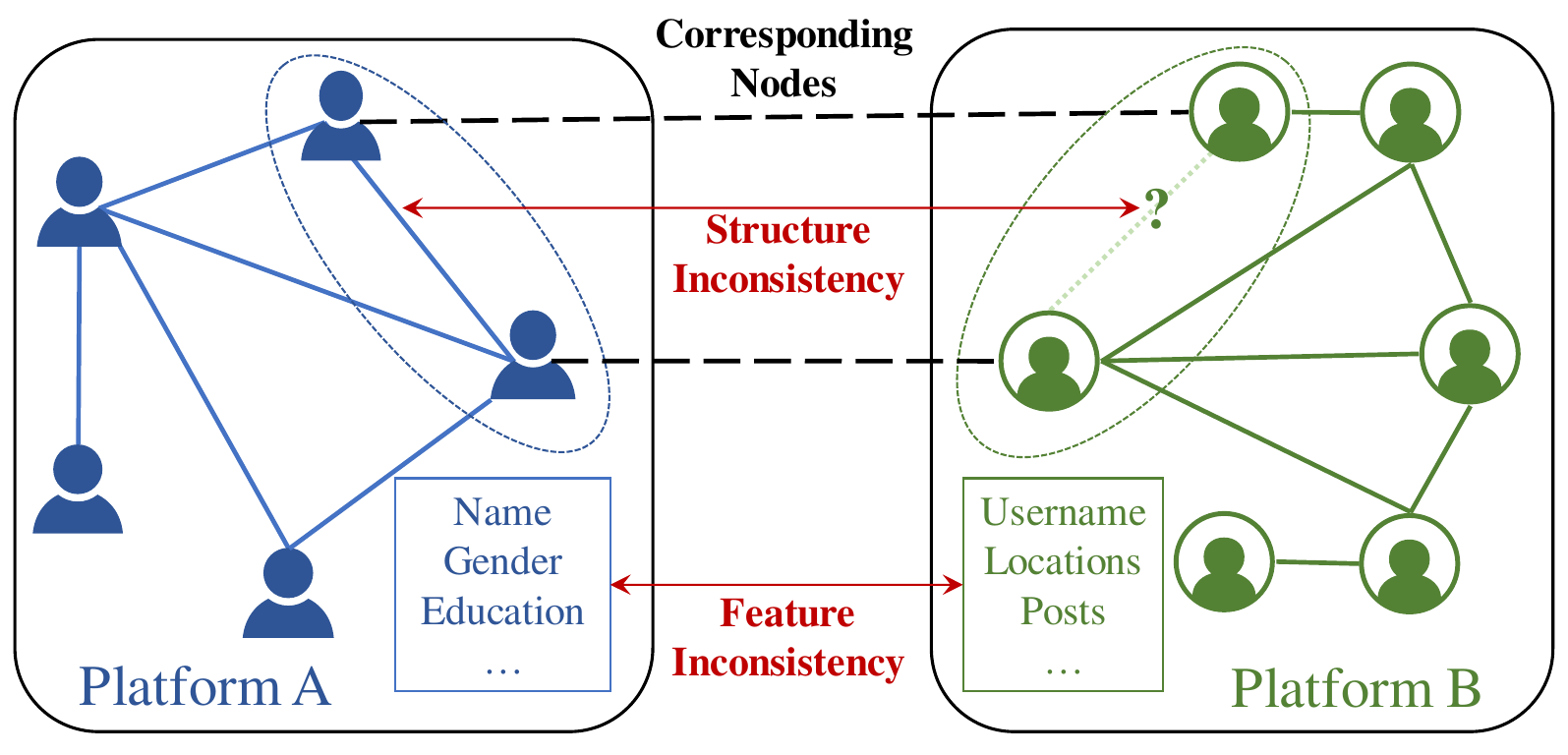} 
	\vspace{-7mm}
	\caption{An example of graph alignment with structure and feature inconsistency.}
	\vspace{-4mm}
	\label{fig:intro}
\end{figure}

Most existing graph alignment methods rely on high-quality and well-measured graph structures. They require that the structure of the overlapped parts between two graphs is similar, which is named \emph{structure consistency} thereafter. However, real-world graphs are often coupled with outliers \cite{tang2022rethinking} or with missing/irrelevant edges \cite{wang2018network,li2021mask,xu2020robust}, leading to structure inconsistency across graphs. It is often observed that the same entities in different networks have quite different neighbors due to the structural noise \cite{chen2019unsupervised, zhou2021unsupervised}. Figure \ref{fig:intro} demonstrates an example of graph alignment on two social network platforms. Black dashed lines are anchor links that connect the same copies of users across two networks. As can be seen, two circled nodes are connected on Platform A, but their corresponding nodes on Platform B are not connected.

Besides structure inconsistency, another largely overlooked issue is that \emph{node features in different graphs are usually unaligned and inconsistent}. Due to various functionalities of different networks (e.g., LinkedIn for job seeking and Twitter for opinion sharing), the same user in different networks commonly does not share the same features. Taking Figure \ref{fig:intro} as an example, user information in Platform A includes the real name, gender, and education experiences, while Platform B contains the anonymized username, locations, posts, etc. Under this situation, corresponding nodes across two networks are not similar to each other and are incomparable. Likewise, in cross-lingual knowledge graph alignment, entities in different languages are usually embedded into individual feature spaces. Using machine translation \cite{mao-etal-2021-alignment, sun2017cross} can alleviate this issue, but may bring additional noise and cost.

\begin{figure}[t!]
	\centering
	\includegraphics[width=0.94\linewidth]{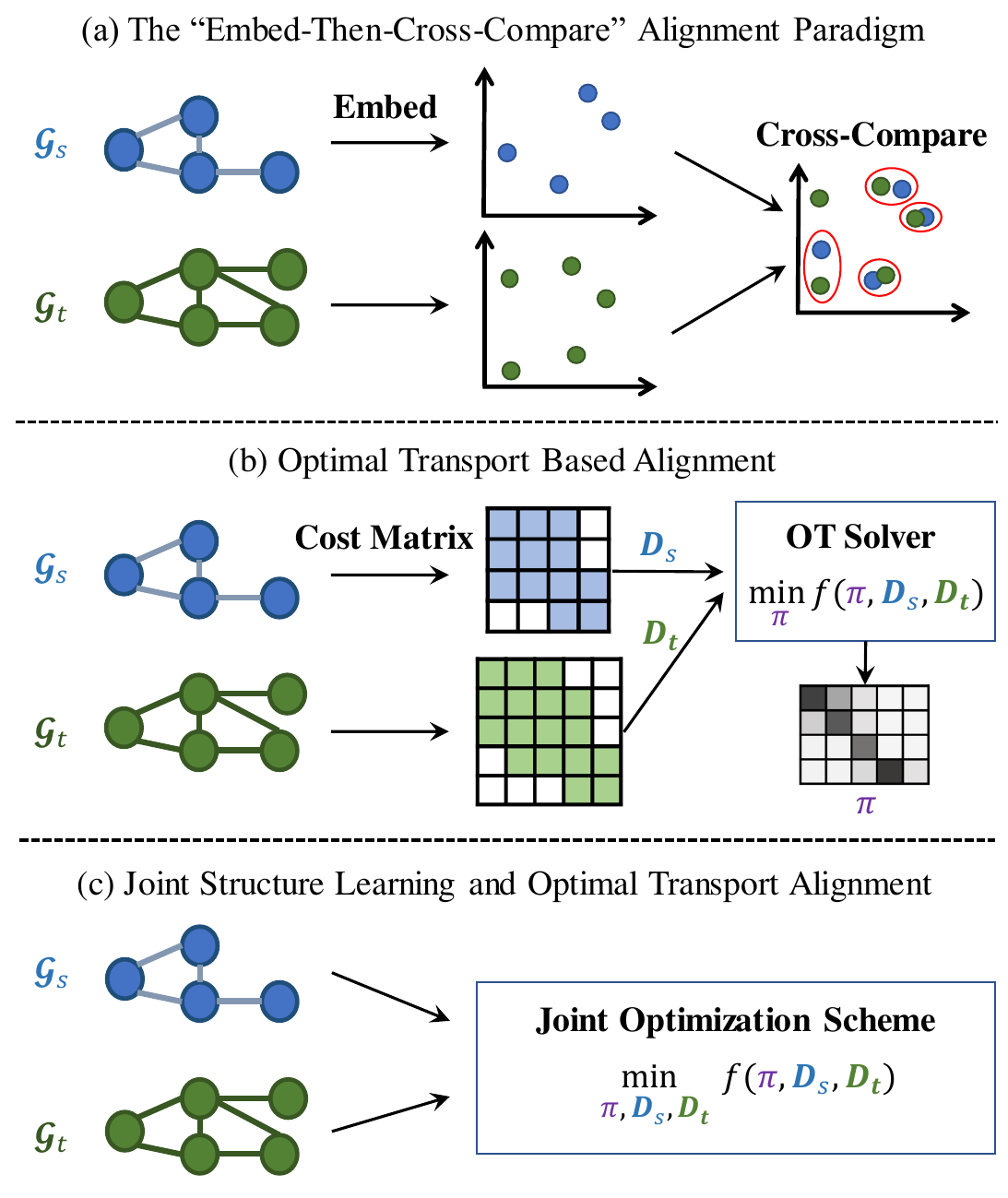} 
	\vspace{-2mm}
	\caption{Comparison between the existing graph alignment methods (top and middle) and our proposed SLOTAlign framework (bottom).}
	\vspace{-2mm}
	\label{fig:intro2}
\end{figure}

In previous works, a popular paradigm for unsupervised graph alignment is the ``embed-then-cross-compare'' procedure \cite{heimann2018regal,chen2019unsupervised,derr2021deep,gao2021unsupervised,liang2021unsupervised,kyster2021boosting}, as shown in Figure \ref{fig:intro2}(a). As the name suggests, it first embeds nodes in each graph into a common feature space (e.g., using a graph neural network), and then compares embeddings across two graphs to obtain node correspondences. Nonetheless, we find this paradigm has the following limitations to deal with structure and feature inconsistency. First, as the node embeddings are calculated by aggregating information from the neighbors, it may amplify noise when structure inconsistency exists. Second, if features in two graphs are inconsistent, the corresponding node embeddings are also typically inconsistent and can not be compared directly \cite{chen2020cone, gao2021unsupervised}.  In knowledge graph alignment, margin-based ranking losses \cite{sun2017cross, sun2020benchmarking} and contrastive learning \cite{zeng2022entity} are frequently used to integrate embedding spaces across graphs. However, without the supervision of ground-truth node pairs, the process of embedding space integration is unstable and unreliable.

Besides the ``embed-then-cross-compare'' paradigm, another line of research is to reformulate the graph alignment problem as finding the optimal probabilistic correspondence between two probability measures on graphs. Specifically, Gromov-Wasserstein (GW) distance serves as an effective tool in modeling the correspondence problems between two graphs on unaligned metric spaces \cite{solomon2016entropic, li2022fast}. We show the procedure of the resulting optimal transport based alignment in Figure \ref{fig:intro2}(b). It first constructs two cost matrices $\bm D_s$ and $\bm D_t$ within each graph. Then, it applies an optimal transport solver to find the best transportation plan $\pi$ with minimal cost according to $\bm D_s$ and $\bm D_t$. The transportation plan $\pi$ reveals node correspondences across graphs.

However, previous optimal transport based methods mainly consider the alignment between plain graphs without attributes and rely on manually designed cost matrices (e.g., the original graph adjacency matrix \cite{xu2019gromov, xu2019scalable, li2022fast} or the heat kernel of graph Laplacian matrix \cite{barbe2020graph, chowdhury2021generalized, li2021deconvolutional}). Thus, these methods are potentially fragile to structure inconsistency. Moreover, how to find  optimal cost matrices in attributed graphs for the optimal transport based alignment has not been well explored by existing methods. In summary, there is no satisfactory solution for enhancing the robustness of graph alignment against structure and feature inconsistency.

To address above issues, we propose a novel framework for joint Structure Learning and Optimal Transport Alignment (SLOTAlign). As shown in Figure \ref{fig:intro2}(c), SLOTAlign simultaneously optimizes the intra-graph structure representation ($\bm D_s, \bm D_t$) and cross-graph transportation plan $\pi$ in a unified manner, which can get rid of choosing cost matrices manually. SLOTAlign models the multi-view structure representation within each graph, which integrates the node-view, edge-view, and subgraph-view to reduce the effect of noise and inconsistency in  original graph structures. Moreover, SLOTAlign is more robust to feature inconsistency as it only utilizes intra-graph node relation and does not depend on cross-graph node embedding comparison. Additionally, we show that SLOTAlign is invariant to graph feature permutation, which cannot be achieved by the ``embed-then-cross-compare'' methods. Theoretically, we  provide an alternating scheme based algorithm to address the optimization problem arisen from SLOTAlign, and establish the convergence result of the proposed algorithm.

To sum up, our contributions are three-fold:
\begin{itemize}[leftmargin=*]
    \item We point out and analyze that existing attributed graph alignment methods are susceptible to both structure and feature inconsistency, and thus perform unstably in noisy real-world graphs.
    \item  A novel framework --- SLOTAlign has been proposed for joint structure learning and optimal transport alignment. We prove the robustness guarantee of SLOTAlign against feature permutation and develop a convergent alternating scheme based algorithm to solve the optimization problem in SLOTAlign. 
    \item We conduct extensive experiments on six graph alignment datasets and the DBP15K KG alignment benchmark dataset. The proposed SLOTAlign shows superior performance over seven general unsupervised graph alignment methods and five specialized KG alignment methods. It also has strongest robustness against multiple types of structure and feature inconsistency.
\end{itemize}

The rest of the paper is organized as follows. Section \ref{sec2} introduces some preliminary knowledge about graph alignment and Gromov-Wasserstein distance. Section \ref{sec3} analyzes the inconsistency issue for graph alignment. Section \ref{sec4} presents the proposed unsupervised graph alignment framework SLOTAlign. Section \ref{sec5} reports the experimental results. Related works and conclusion are presented in Section \ref{sec6} and Section \ref{sec7} respectively.

%% file: sec2_preliminary.tex
\section{Preliminary}\label{sec2}

We denote an undirected attributed graph as $\mathcal G=\{\mathcal V, \bm A, \bm X\}$, where $\mathcal V=\{v_1, v_2, \cdots, v_n\}$ is the set of $n$ nodes, $\bm A$ is the adjacency matrix of the graph. $\bm A_{ij}$ equals to 1 if there is an unweighted edge between $v_i$ and $v_j$, otherwise 0. $\bm X \in \mathbb R^{n\times d}$ is the node feature matrix. $\textbf x_i=\bm X(i,:)\in \mathbb R^{d}$ is the feature vector of node $v_i$. 

\subsection{Problem Statement}

In this paper, we consider the problem of aligning two attributed graphs in an unsupervised manner. We refer to one graph as the source graph and the other as the target graph, denoted with $\mathcal G_s$ and $\mathcal G_t$ respectively. For each node in the source graph, graph alignment aims to identify, if any, the corresponding node in the target graph. Moreover, unsupervised alignment methods do not require any ground-truth node correspondences. We formulate this problem as follows.

\begin{definition}[Unsupervised Attributed Graph Alignment] \label{def:task}
Given two attributed graphs $\mathcal G_s=(\mathcal U_s, \bm A_s, \bm X_s)$ and $\mathcal G_t=(\mathcal V_t, \bm A_t, \bm X_t)$, without any observed node correspondences, the unsupervised graph alignment algorithm returns a set of aligned node pairs $\mathcal M=\{(u_i,v_j)|(u_i,v_j)\in \mathcal U_s \times \mathcal V_t \}$, where $u_i$ and $v_j$ are corresponding nodes across two graphs. 
\end{definition}

\subsection{Gromov-Wasserstein Distance for Alignment}
Conventional optimal transport needs a ground cost $C$ to compare probability measures $(\mu,\nu)$ and thus cannot be used if the measures are not defined on the same underlying space \cite{peyre2019computational}. To address this limitation, the Gromov-Wasserstein (GW) distance was originally proposed in \cite{memoli2011gromov} for quantifying the distance between two probability measures supported on unaligned metric spaces. More precisely:
\begin{definition}[GW Distance]
    \label{def:gw}
    Suppose that we are given two unregistered compact metric spaces $(\mathcal{X},d_X)$, $(\mathcal{Y},d_Y)$ accompanied with Borel probability measures $\mu,\nu$ respectively. The GW distance between $\mu$ and $\nu$ is defined as 
    \[
    \inf_{\pi \in \Pi(\mu,\nu)}  \iint |d_X(x,x')-d_Y(y,y')|^2 d\pi(x,y)d\pi(x',y'),
    \]
    where $\Pi(\mu,\nu)$ is the set of all probability measures on $\mathcal{X}\times\mathcal{Y}$ with $\mu$ and $\nu$ as marginals. 
 \end{definition}
Intuitively, the GW distance is trying to preserve the isometric structure between two probability measures under the optimal derivation. If a map pairs $x\rightarrow y$ and $x'\rightarrow y'$, then the distance between $x$ and $x'$ is supposed to be close to the distance between $y$ and $y'$. Notably, the GW distance only requires modeling the topological or relational aspects of the distributions within each domain. In view of these nice properties, the GW distance has attracted intense research over the last decade, especially for structured data analysis, e.g., molecule analysis~\cite{vayer2018fused,titouan2019optimal}, 3D shape matching~\cite{solomon2016entropic, li2022fast}, graph embedding and classification~\cite{vayer2019sliced,vincent2021online}, generative models~\cite{bunne2019learning,xu2021learning}, to name a few.

To apply the GW distance on the graph alignment problem, we consider the discrete case of Definition \ref{def:gw}. Suppose $\mu$ is a uniform distribution over all $n$ nodes in the source graph $\mathcal G_s$ and $\nu$ is a uniform distribution over all $m$ nodes in the target graph $\mathcal G_t$, i.e., $\mu = \frac{1}{n}\sum_{i=1}^n \delta_{u_i}$ and $\nu = \frac{1}{m}\sum_{j=1}^m \delta_{v_j}$, where $\delta_{u_i}$ and $\delta_{v_j}$ are one-hot signals on node $u_i$ and $v_j$. The GW distance between $\mu$ and $\nu$ can be reformulated as:
\begin{equation}
\begin{aligned}
\label{eq:graph_gw}
&\min_{ \pi \in \Pi(\mu,\nu)} \sum_{i=1}^n \sum_{j=1}^n \sum_{k=1}^m \sum_{l=1}^m |\bm D_s(i,j)-\bm D_t(k,l)|^2 \pi_{ik}\pi_{jl},\\
& \quad \,\text{s.t.} \quad\pi \mathbf{1}_m = \mu,\,\,  \pi^T\mathbf{1}_n = \nu,\,\,   \pi \ge 0,\\
\end{aligned}
\end{equation}
where $\bm D_s(i,j)$ can be considered as the transportation cost between $u_i$ and $u_j$ in $\mathcal G_s$ (e.g., the edge $\bm A_{ij}$), and $\bm D_t(k,l)$ is the cost between $v_k$ and $v_l$ in $\mathcal G_t$. In \eqref{eq:graph_gw}, if $\pi_{ik}$ and $\pi_{jl}$ are large which indicates ($u_i$, $v_k$) and ($u_j$, $v_l$) are likely to be two node pairs, the difference of the corresponding intra-graph transportation costs should be similar, i.e., $|\bm D_s(i,j)-\bm D_t(k,l)|\to 0$.

Accordingly, the GW distance optimization problem solves the optimal transport $\pi$ merely based on two intra-graph cost matrices $\bm D_s$ and $\bm D_t$. In graph alignment, $\pi_{ik}$ indicates the matching score between $u_i$ in $\mathcal G_s$ and $v_k$ in $\mathcal G_t$, and the alignment $\mathcal M$ can be derived from $\pi$:
\begin{equation}\label{eq:match}
\mathcal M =  \mathop{\arg\max}_{\mathcal M \in \mathbb M} \sum_{(u_i, v_k) \in \mathcal M} \pi_{ik},
\end{equation}
where $\mathbb M$ is the set of all legit alignments.

%% file: sec3_analysis.tex
\section{Analysis of the Inconsistency Issue}\label{sec3}

In this section, we discuss in-depth why structure inconsistency and feature inconsistency are challenging issues in the task of unsupervised graph alignment. As we mentioned before, a popular paradigm for solving graph alignment is first embedding nodes in two graphs as well as possible and second converting it to a cross-graph computation problem in the embedding space, i.e., an ``embed-then-cross-compare'' procedure. More specifically, it first calculates node embeddings in each graph, for example, using a graph neural network. Then, it computes a set of node pairs $\mathcal M$ based on the closeness of node embeddings across two graphs by solving the following embedding matching problem: 
\begin{equation}\label{eq:analysis1}
    \min_{\mathcal M \in \mathbb M} \sum_{(u_i, v_j) \in \mathcal M} \|\mathbf z_{u_i} - \mathbf z_{v_j}\|,
\end{equation}
where $\mathbf z_{u_i}$ and $\mathbf z_{v_j}$ denote the embeddings of node $u_i$ in $\mathcal G_s$ and node $v_j$ in $\mathcal G_t$, respectively.

Nonetheless, this ``embed-then-cross-compare'' paradigm requires that $\mathbf z_{u_i}$ and $\mathbf z_{v_j}$ are in the same embedding space. When structures and features across two graphs are inconsistent, the corresponding node embeddings $\mathbf z_{u_i}$ and $\mathbf z_{v_j}$ are also typically unaligned and incomparable \cite{chen2020cone,gao2021unsupervised}. Before matching node embeddings across graphs, we need to align the embedding spaces of two graphs, e.g., using a linear transformation $\bm Q$: 
\begin{equation*}
    \min_{\mathcal M \in \mathbb M} \min_{\bm Q} \sum_{(u_i, v_j) \in \mathcal M} \|\mathbf z_{u_i} \bm Q  - \mathbf z_{v_j}\|.
\end{equation*}
In the unsupervised graph alignment task, as none of the node correspondence is known, $\bm Q$ is difficult to be calculated directly. On the other hand, because $\bm Q$ is unknown, we are unable to compare node embeddings across graphs to obtain $\mathcal M$. To resolve this ``chicken-and-egg'' paradox, the adversarial learning framework is adopted \cite{gao2021unsupervised, chen2019unsupervised,zhou2021unsupervised} to generate pseudo node correspondences or auxiliary learning signals. However, these methods still require two embedding spaces to be partially aligned, otherwise the stability of iterations is not guaranteed.

On another front, optimal transport based methods that use Gromov-Wasserstein distance for graph alignment are more resilient to feature inconsistency as they mainly consider the intra-graph structural information. But as the price, they may be more fragile to the inconsistency in graph structures.

To better illustrate the robustness issue of existing methods under structure and feature inconsistency. We use the Cora citation network \cite{sen2008collective, yan2021bright} with the first 100 feature columns as the source graph $\mathcal G_s$ and generate the target graph $\mathcal G_t$ by producing different levels of inconsistency in $\mathcal G_s$. To control structure inconsistency, we randomly perturb $p\%$ edges in $\mathcal G_t$ to other previous unconnected positions and keep node features unchanged. For feature inconsistency, we fix the edge perturbation ratio to $25\%$ and randomly permute the order of $p\%$ feature columns in $\mathcal G_t$.

\begin{figure}[t!]
	\centering
	\includegraphics[width=\linewidth]{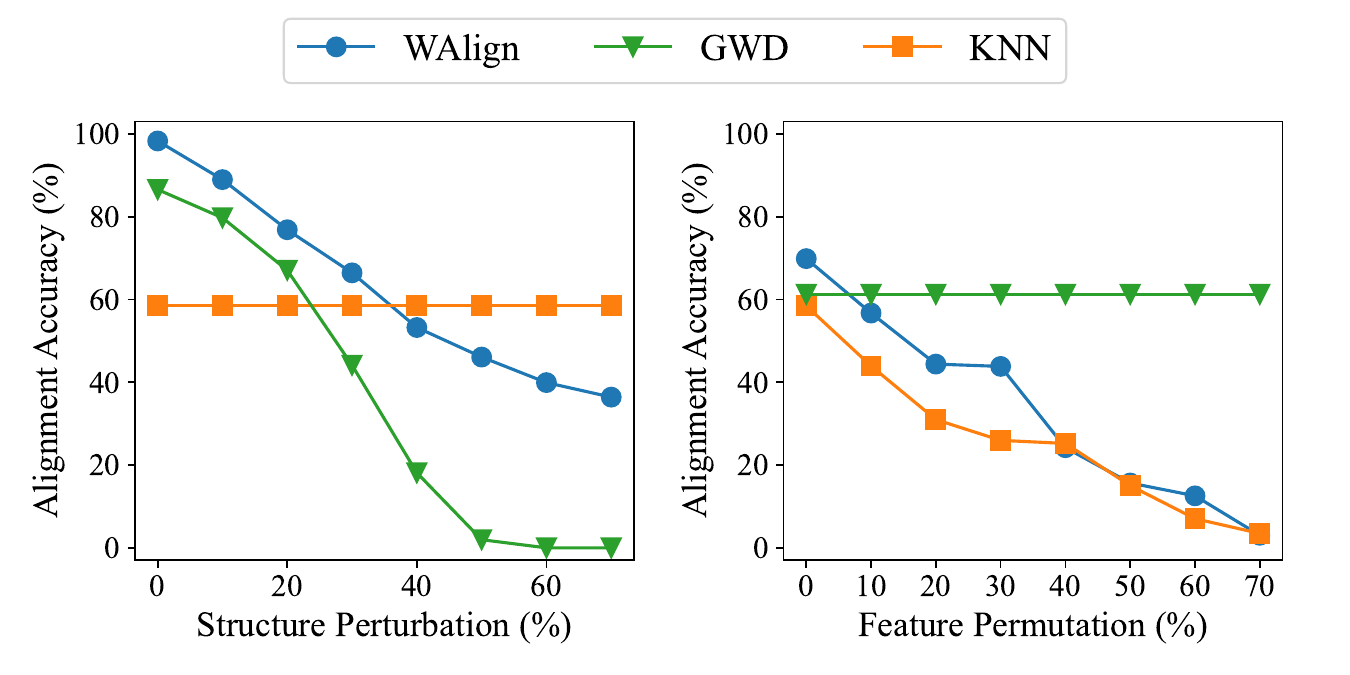} 
	\vspace{-7mm}
	\caption{Performance comparison of three graph alignment methods under different level of structure and feature inconsistency.}
	\vspace{-4mm}
	\label{fig:analysis}
\end{figure}

We compare the alignment performance of \textbf{WAlign} \cite{gao2021unsupervised}, the start-of-the-art unsupervised graph alignment method based on adversarial training, with \textbf{GWD} \cite{xu2019gromov}, a GW distance based method using graph adjacency matrices as cost matrices for alignment, and K-Nearest Neighbor (\textbf{KNN}), a simple baseline that directly matches nodes according to feature similarity. In Figure \ref{fig:analysis}, we observe that the performance of WAlign is significantly affected by both structure and feature inconsistency. For example, when structure perturbation ratio is larger than 40\%, WAlign is beat by KNN. Likewise, when feature permutation ratio is greater than 40\%, the performance of WAlign is also very close to KNN. As for GWD, it is not influenced by any degree of feature inconsistency, but is more vulnerable to structure inconsistency.

Due to the above robustness issues in previous methods, in this work, we approach the graph alignment task via intra-graph structure modeling and cross-graph optimal transport alignment in a unified manner. We take the advantages of both embedding-based methods and optimal transport based methods, and show that it is more robust against structure and feature inconsistency. We introduce the proposed approach in the next section.

%% file: sec4_method.tex
\begin{figure*}[tb]
	\centering
	\includegraphics[width=0.9\textwidth]{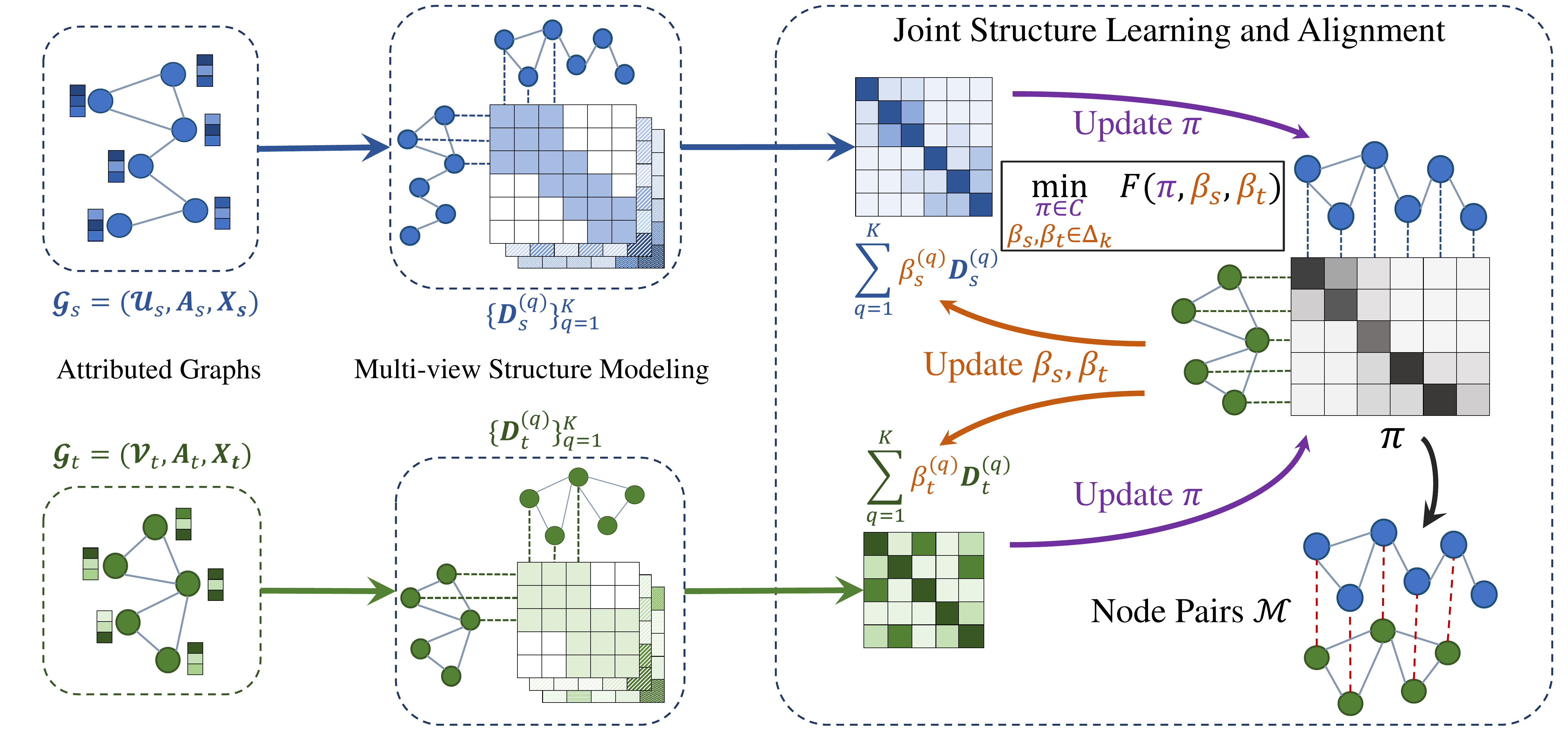} 
	\vspace{-2mm}
	\caption{The framework of SLOTAlign. Given two attribute graphs, it first constructs a set of candidate structure views for each graph, and then optimizes the integrated structure representations and node correspondences simultaneously in the joint structure learning and graph alignment framework.}
	\vspace{-2mm}
	\label{fig:model}
\end{figure*}

\section{Methodology}\label{sec4}
We propose a unified framework, SLOTAlign, that jointly performs Structure Learning and Optimal Transport Alignment. As introduced in section \ref{sec2}, the Gromov-Wasserstein (GW) distance is able to establish a connection between two graphs on unaligned metric spaces.  With the help of the GW distance, the graph alignment task can be regarded as an optimal transport problem between two intra-graph matrices without the requirement of cross-graph comparison. However, most existing optimal transport based methods only consider the alignment between plain graphs and rely on a fundamental assumption that original graph structure can be viewed as ground-truth information for alignment. Unfortunately, such assumption is usually violated in real-world scenarios as we discussed in Section \ref{sec1} and \ref{sec3}. Although the Fused GW Distance \cite{titouan2019optimal} attempts to take node attributes into consideration, the cost matrices are still manually constructed which are fragile to the structure and feature inconsistency in  real-world graphs.

To navigate such a pitfall, we are motivated to learn an optimal representation of the graph structure for alignment. We design a multi-view structure representation, including the node-view, edge-view, and subgraph-view, to model different perspectives of the original graph. We then develop a joint structure learning and alignment framework, which finds the optimal structure representation and node correspondences simultaneously. Figure \ref{fig:model} illustrates the framework of SLOTAlign.

\begin{figure}[t]
	\centering
	\includegraphics[width=\linewidth]{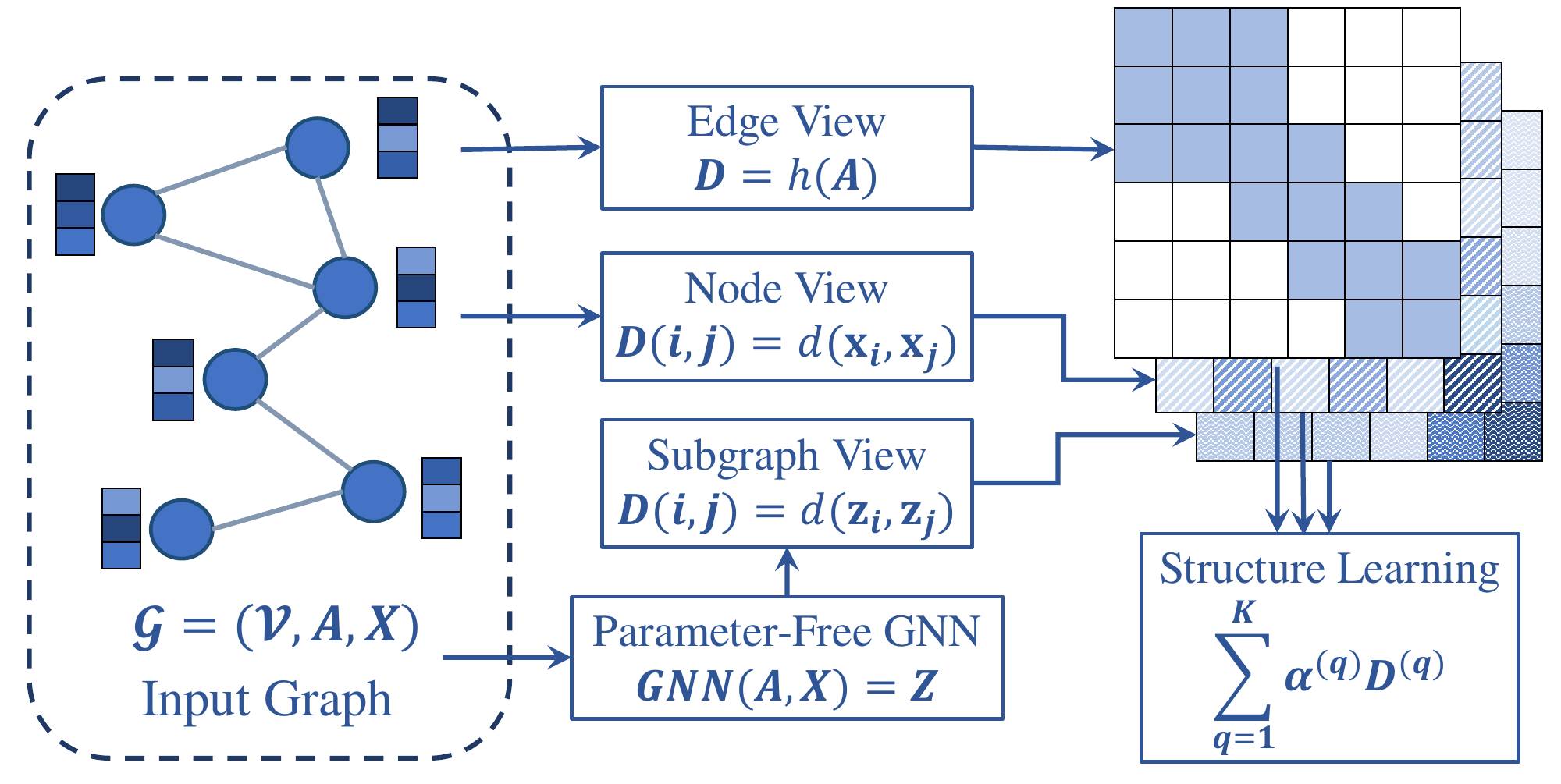} 
	\vspace{-4mm}
	\caption{The illustration of multi-view structure learning.}
	\vspace{-3mm}
	\label{fig:structure}
\end{figure}

\subsection{Multi-view Structure Modeling}\label{sec41}

Based on Equation \eqref{eq:graph_gw}, the optimization objective of the GW distance can be rewritten by
\begin{equation}\label{eq:model1}
    \begin{aligned}
    F(\pi)  = & \frac{1}{n^2}\sum_{i=1}^n \sum_{j=1}^n \bm D_s(i,j) ^2 + \frac{1}{m^2}\sum_{k=1}^m \sum_{l=1}^m \bm D_t(k,l) ^2 \\
    & -2\textnormal{tr}(\bm D_s\pi \bm D_t\pi^T). 
    \end{aligned}
\end{equation}

It is not hard to observe that the alignment quality from the GW distance heavily depends on how to construct two intra-graph cost matrices $\bm D_s$ and $\bm D_t$. Instead of designing $\bm D_s$ and $\bm D_t$ manually, we first construct two sets of candidate graph structure bases $\{\bm D^{(q)}_s\}_{q=1}^K$ and $\{\bm D^{(q)}_t\}_{q=1}^K$ that enhance the original structure from multiple different views. As shown in Figure \ref{fig:structure}, we consider the following perspectives in multi-view structure modeling:

\textbf{(1) Edge-view.} It models the new graph structure as a function of the original adjacency matrix, i.e., $\bm D=h(\bm A)$. Here $h:\mathbb R^{n\times n}\to \mathbb R^{n\times n}$ can be any legit matrix mapping function such as the power up operator $h(\bm A)=\bm A^k$. For simplicity and efficiency in practice, we use the identity mapping: $\bm D=h(\bm A)=\bm A$.

\textbf{(2) Node-view}. It uses pairwise node feature similarity to represent the new graph structure, i.e., $\bm D(i,j)=d(\textbf x_i, \textbf x_j)$ where $\textbf x_i$ and $\textbf x_j$ are the feature vectors of node $v_i$ and $v_j$, and $d$ can be any similarity metric such as cosine similarity. Here we use the inner product to measure node similarity: $\bm D(i,j)=d(\textbf x_i, \textbf x_j)=\textbf x_i^T \textbf x_j$, which is equivalent to cosine similarity after nodewise feature normalization.

Both the node-view and the edge-view consider the first-order relation between nodes. To model high-order interaction on the graph, we further construct the subgraph-view:

\textbf{(3) Subgraph-view}. It first integrates node features and neighbor information in the original structure using a Graph Neural Network (GNN), and then calculates the pairwise similarity between node embeddings. However, the training process of GNN may not be stable in unsupervised graph alignment, as discussed in Section \ref{sec3}. Thus, we adopt a parameter-free GNN derived from the simplified graph convolutional network\cite{wu2019simplifying}. Specifically, we remove the parameterized linear layer and the activation function:
\begin{equation}
    \mathbf Z^{(k)}=\hat{\bm A}^k\bm X=(\bm M^{-\frac{1}{2}}(\bm A+\bm I)\bm M^{-\frac{1}{2}})^k\bm X
\end{equation}
where $\bm I$ is the identity matrix, $\bm M$ is the degree matrix of $\bm A+\bm I$, and $\hat{\bm A}=\bm M^{-\frac{1}{2}}(\bm A+\bm I)\bm M^{-\frac{1}{2}}$ is the symmetric normalized adjacency matrix after adding self-loops for all nodes. After feature propagation for $k$ steps, the node embedding $\bm Z^{(k)}$ contains the $k$-hop neighborhood subgraph information of each node. Suppose $\mathbf z_i^{(k)}$ is the embedding of node $v_i$, we also use the inner product to measure the $k$-order sub-graph similarity between nodes $v_i$ and $v_j$: $\bm D(i,j)=d(\mathbf z^{(k)}_i, \mathbf z^{(k)}_j)={\mathbf z_i^{(k)}}^T \mathbf z^{(k)}_j,\,k=1,2,3,\cdots$.

To sum up, suppose $K\geq 3$ is the maximum number of candidate structure bases, SLOTAlign construct the following bases $\{\bm D^{(q)}_s\}_{q=1}^K$ for $\mathcal G_s$:
\begin{equation}
\small
\begin{aligned}
\label{eq:proximity}
    &\bm D^{(1)}_s=\bm A_s ~(\textnormal{edge-view}), \ \ \ \ \bm D^{(2)}_s=\bm X_s\bm X_s^T ~(\textnormal{node-view}),\\
    &\bm D^{(q)}_s=\hat{\bm A}_s^{q-2} \bm X_s (\hat{\bm A}_s^{q-2} \bm X_s)^T, 2<q\leq K~(\textnormal{subgraph-view}).\\
\end{aligned}
\end{equation}
For $\mathcal G_t$, the construction process of $\{\bm D^{(q)}_t\}_{q=1}^K$ is the same.

\subsection{Joint Structure Learning and Optimal Transport Alignment}\label{sec42}

After we construct  multi-view graph structure bases  $\{\bm D^{(q)}_s\}_{q=1}^K$ and $\{\bm D^{(q)}_t\}_{q=1}^K$, a natural question is how to integrate these candidate bases and learn an optimal representation of the graph structure. To this end, we model the new graph structure $\bm D_s$ and $\bm D_t$ as a convex hull of candidate structure bases. Specifically, we define
\begin{equation}
    \bm D_s = \sum_{q=1}^K \beta^{(q)}_s \bm D^{(q)}_s,~~~ \bm D_t = \sum_{q=1}^K \beta^{(q)}_t \bm D^{(q)}_t,
\end{equation}
where $\beta_s, \beta_t \in \Delta_K$, i.e., $\sum_{q=1}^K \beta^{(q)}_s = 1, \beta^{(q)}_s \ge 0, \forall q \in [K]$.

Subsequently, we target to optimize the weight vectors $\beta_s, \beta_t$ and the transportation policy $\pi$ in a unified Gromov-Wasserstein framework. More specifically, Equation \eqref{eq:model1} can be recast into the resulting optimization problem, i.e., 
\begin{equation}
\label{eq:main_opt}
    \begin{aligned}
    \min_{\pi\in C, \beta_s, \beta_t \in \Delta_K} F(\pi, \beta_s, \beta_t),
    \end{aligned}
\end{equation}
where $C=\{\pi\ge0: \pi\mathbf{1}_m = \mu, \pi^T\mathbf{1}_n = \nu\}$ and $F(\cdot,\cdot)$ is a bi-quadratic function, i.e.,  
\begin{equation}\label{eq:main_opt2}
\small
    \begin{aligned}
    F(\pi,\beta_s,\beta_t) &= \frac{1}{n^2}\sum_{i=1}^n \sum_{j=1}^n  \left(\sum_{q=1}^K \beta_s^{(q)} \bm D_s^{(q)}(i,j) \right)^2 \\
    & + \frac{1}{m^2}\sum_{k=1}^m \sum_{l=1}^m  \left(\sum_{q=1}^K \beta_t^{(q)} \bm D_t^{(q)}(k,l) \right)^2 \\
    & -2\textnormal{tr}\left(\left(\sum_{q=1}^K \beta_s^{(q)} \bm D_s^{(q)}\right)\pi \left(\sum_{q=1}^K \beta_t^{(q)} \bm D_t^{(q)}\right)\pi^T\right). 
    \end{aligned}
\end{equation}
Instead of purely finding the optimal probabilistic correspondence relationship in the vanilla GW problem, the proposed SLOTAlign is trying to learn the optimal structure representation and matching relationship simultaneously. To better understand the proposed SLOTAlign, we can consider a simple case --- the weight vectors are vertices of the simplex, e.g., $\beta_s = \beta_t = (1,0,\cdots,0)$. Then, \eqref{eq:main_opt} will reduce to the vanilla model studied in previous work \cite{xu2019gromov}. As such, if we minimize the GW objective on both $\beta_s, \beta_t$ and $\pi$, the optimal solution returned from \eqref{eq:main_opt} will be better than the original one.

Next, we shed light on the robustness of SLOTAlign against feature inconsistency by a specific case --- feature permutation. 

\begin{definition}[Permutation on graph features]\label{def:PI}
  Let $\bm P \in \{0,1\}^{d \times d}$ be any permutation matrix of order $d$, the feature permutation $\mathcal P$ on a graph $\mathcal G = \{\mathcal V, \bm A, \bm X\}$ is defined as a mapping of the node feature indices, i.e., $  \mathcal P(\mathcal G) = \{\mathcal V, \bm A, \bm X\bm P\}$. 
\end{definition}

Feature permutation is very common in real-world cases, which represents two graphs sharing the same feature types but in different orders. Notably, methods based on ``embed-then-cross-compare'' are unstable to feature permutation as the feature space in $\mathcal G_t$ is changed and no more aligned with $\mathcal G_s$. In comparison, SLOTAlign would not be affected by feature permutation according to the following proposition.

\begin{proposition}\label{prop:1}
\textnormal{SLOTAlign} is invariant to feature permutation $\mathcal P$ on $\mathcal G_s$ or $\mathcal G_t$, e.g., $\textnormal{SLOTAlign}(\mathcal G_s, \mathcal G_t) = \textnormal{SLOTAlign}(\mathcal G_s, \mathcal P(\mathcal G_t))$
\end{proposition}

Assume that the structure bases of the permuted graph $\mathcal P(\mathcal G_t)=\{\mathcal V, \bm A_t, \bm X_t\bm P\}$ are $\{\overline{\bm D}^{(q)}_t\}_{q=1}^K$, we have 
\begin{equation}
\small
\begin{aligned}
    &\overline{\bm D}^{(1)}_t = \bm D^{(1)}_t,\quad \overline{\bm D}^{(2)}_t=\bm X_t\bm P\bm P^T\bm X_t^T=\bm X_t\bm X_t^T=\bm D^{(2)}_t,\\
    &\overline{\bm D}^{(q)}_t=\hat{\bm A}_t^{q-2} \bm X_t\bm P \bm P^T \bm X_t^T (\hat{\bm A}_s^{q-2})^T = \bm D^{(q)}_t, \ \  2<q\leq K.\\
\end{aligned}
\end{equation}
Therefore, we have $\{\overline{\bm D}^{(q)}_t\}_{q=1}^K=\{\bm D^{(q)}_t\}_{q=1}^K$. As $\mathcal P(\mathcal G_t)$ and $\mathcal G_t$  have the same structure bases, the optimization problem \eqref{eq:main_opt2} in SLOTAlign is unchanged, and we complete the proof. The robustness of the proposed SLOTAlign against more types of inconsistency has been further demonstrated in the experiment section.

\subsection{Optimization Algorithm}\label{sec43}

\begin{algorithm}[t]
\small
\SetAlgoLined
\KwIn{1. Source graph $\mathcal G_s=(\mathcal U_s, \bm A_s, \bm X_s)$\\
\qquad \ \ \ 2. Target graph $\mathcal G_t=(\mathcal V_t, \bm A_t, \bm X_t)$\\
\qquad \ \ \ 3. Maximum number of iterations $k_{max}$\\
\qquad \ \ \ 4. Number of candidate structure bases $K$\\
\qquad \ \ \ 5. Step size of structure learning $\tau$\\
\qquad \ \ \ 6. Step size in the Sinkhorn algorithm $\eta$\\
}
\KwOut{Set of node correspondence pairs $\mathcal{M}$}
Initialize $\beta^{(q)}_s \gets \frac{1}{K}$, $\beta^{(q)}_t \gets \frac{1}{K} (1\leq q \leq K)$;\\
Initialize $\alpha^1 \gets [\beta_s,\beta_t]$;\\
Initialize $\pi^1_{ij} \gets \frac{1}{nm} (1\leq i \leq n, 1\leq j \leq m)$;\\
Construct candidate structure bases $\{\bm D^{(q)}_s\}_{q=1}^K$ and $\{\bm D^{(q)}_t\}_{q=1}^K$ according to Equation \eqref{eq:proximity};\\
\For{$k=1...k_{max}$}{
    Update $\alpha^{k+1} \gets \alpha^{k}$ according to Equation \eqref{eq:alpha_update};\\
    Update $\pi^{k+1} \gets \pi^{k}$ according to Equation \eqref{eq:pi_update};\\
    \If{$|\alpha^{k+1} - \alpha^k| < \epsilon_1$ \textnormal{and} $|\pi^{k+1} - \pi^k| < \epsilon_2$}{
        Break;
    }
}
Generate node pairs $\mathcal M$ according to Equation \eqref{eq:match};
\caption{SLOTAlign}\label{alg}
\end{algorithm}

In this subsection, we provide a theoretically sound optimization algorithm to tackle \eqref{eq:main_opt}. To proceed, we detect the hidden structure of \eqref{eq:main_opt} at first and further take it account into the algorithmic development. We can observe that \eqref{eq:main_opt} is a nonconvex bi-quadratic program with polytope constraints. The basic strategy here is to optimize the weight $\beta_s, \beta_t$ and the matching matrix $\pi$ in an alternating fashion. As $\beta_s$ and $\beta_t$ are regarded as one block in our algorithm design, for simplicity, we use $\alpha = [\beta_s,\beta_t]$ to represent the concatenation of $\beta_s$ and $\beta_t$, i.e., $F(\pi, \alpha)=F(\pi, \beta_s, \beta_t)$. More specifically, we  adopt the proximal alternating linearized minimization strategy  \cite{bolte2014proximal} here. To start with, we focus on the 
$\alpha$ update: 
\begin{equation}
\label{eq:alpha_update}
\begin{aligned}
    \alpha^{k+1}  & = \mathop{\arg\min}_{\alpha \in \Theta} \left\{ \nabla_\alpha F(\pi^k,\alpha^k)^T\alpha + \frac{1}{2\tau}\|\alpha-\alpha^k\|^2\right\} \\
    &  = \textnormal{Proj}_{\Theta}\left(\alpha^k-\tau\nabla_\alpha F(\pi^k,\alpha^k) \right).  
 \end{aligned}
\end{equation}
Here, $\Theta =\{\alpha:\sum_{q=1}^K \alpha_q = \sum_{q=K+1}^{2K}\alpha_q = 1, \alpha_q \ge 0, \forall q \in [2K]\}$. Due to the separable structure over $\beta_s$ and $\beta_t$, \eqref{eq:alpha_update} can be reduced to two simplex projection problems, 
 which can be solved efficiently \cite{duchi2008efficient}. 

The crux of our algorithm is the $\pi$-update. As the projection onto the Birkhoff polytope (i.e., $C$) is rather computationally demanding, we are motivated to apply the entropic regularization to handle the $\pi$-update. 
Instead of the Euclidean distance $\|\cdot\|^2$, we invoke the Kullback-Leibler divergence. As such, the $\pi$-update is identical to the entropic optimal transport problem, and thus we can invoke the Sinkhorn algorithm to tackle it efficiently~\cite{cuturi2013sinkhorn}. 
\begin{equation}
\label{eq:pi_update}
    \pi^{k+1}  = \mathop{\arg\min}_{\pi\in C}\left\{\nabla_\pi F(\pi^k,\alpha^{k+1})^T\pi + \frac{1}{\eta} \textbf{KL}(\pi||\pi^k)\right\},
\end{equation}
where $\textbf{KL}(\cdot||\cdot)$ is the Kullback-Leibler divergence, i.e., 
\[\mathbf{K L}(\mathbf{P} 
|| \mathbf{K}) \stackrel{\text { def. }}{=} \sum_{i, j} \mathbf{P}_{i, j} \log \left(\frac{\mathbf{P}_{i, j}}{\mathbf{K}_{i, j}}\right)-\mathbf{P}_{i, j}+\mathbf{K}_{i, j}.\]

We illustrate the detailed initialization and iteration process of SLOTAlign in Algorithm \ref{alg}. Given two sets of intra-graph structure bases, SLOTAlign solves the optimization problem \eqref{eq:main_opt} by updating $\alpha$ and $\pi$ alternatively via Equation \eqref{eq:alpha_update} and \eqref{eq:pi_update}.

A further natural question is whether the proposed algorithm will converge or not. We answer the question in the affirmative. 
\begin{theorem}
\label{thm}
Suppose that  $0<\tau<\tfrac{1}{L_f^\alpha}$ and $0<\epsilon<\eta<\tfrac{1}{L_f^\pi}$, where $L_f^\pi$ and $L_f^\alpha$ are the gradient Lipschitz continuous modulus of $F(\pi,\alpha)$ respectively. Then,
any limit point of the sequence $\{(\pi^{k},\alpha^k)\}_{k \ge 0 }$ converges to a critical point of $\overline{F}(\pi, \alpha)$, i.e., 
\[\overline{F}(\pi, \alpha) = F(\pi, \alpha) + \mathbf{I}_{C}(\pi) + \mathbf{I}_{ \Theta}(\alpha),\]
where $\mathbf{I}_C(\cdot)$ is the so-called indicator function on the set $C$. 
\end{theorem}
\begin{proof}
 As both $C$ and $\Theta$ are bounded sets, then $\{(\pi^k,\alpha^k)\}$ is a bounded sequence. Subsequently, we target at proving the sufficient decrease property. As ${F}(\cdot,\cdot)$ is a bi-quadratic function and the sequence $\{(\pi^{k}, \alpha^k)\}_{k \ge 0}$, $F(\pi,\alpha)$ is gradient Lipschitz continuous with modulus $L_f^\pi$ and $L_f^\alpha$. To proceed, we revisit the $\alpha$-update at first,  
 \[
 \alpha^{k+1}  = \mathop{\arg\min}_{\alpha \in \Theta} \left\{ \nabla_\alpha F(\pi^k,\alpha^k)^T\alpha + \frac{1}{2\tau}\|\alpha-\alpha^k\|^2\right\}.
 \]
 Since $\alpha^{k+1}$ is the optimal solution of a strongly convex problem, we have 
 \begin{align*}
     & \nabla_\alpha F(\pi^k,\alpha^k)^T\alpha^k + \mathbf{I}_{\Theta}(\alpha^k) \ge \\ & \nabla_\alpha F(\pi^k,\alpha^k)^T\alpha^{k+1} + \mathbf{I}_{\Theta}(\alpha^{k+1}) +  \frac{1}{2\tau}\|\alpha^{k+1}-\alpha^k\|^2. 
 \end{align*}
Based on the gradient Lipschitz continuous property, it is easy to obtain, 
\begin{align*}
& F(\pi^k,\alpha^k)+\mathbf{I}_{\Theta}(\alpha^k)-F(\pi^k,\alpha^{k+1})-\mathbf{I}_{\Theta}(\alpha^{k+1}) \\
\ge &\left(\frac{1}{2\tau}-\frac{L_f^\alpha}{2}\right)\|\alpha^{k+1}-\alpha^k\|^2.
\end{align*}
On another front, we recall the $\pi$-update: 
 \[
 \pi^{k+1}  = \mathop{\arg\min}_{\pi\in C}\left\{\nabla_\pi F(\pi^k,\alpha^{k+1})^T\pi + \frac{1}{\eta} \textbf{KL}(\pi||\pi^k)\right\}.
\]
Notably, $\textbf{KL}(\pi||\pi^k)$ is 1-strongly convex on the Birkhoff polytope constraint, i.e., 
\[
\textbf{KL}(\pi^{k+1}||\pi^k) \ge \frac{1}{2}\|\pi^{k+1}-\pi^k\|_F^2. 
\]
Similarly, we have 
\begin{align*}
& F(\pi^k,\alpha^{k+1})+\mathbf{I}_{C}(\pi^k)-F(\pi^{k+1},\alpha^{k+1})-\mathbf{I}_{C}(\pi^{k+1}) \\
\ge &\left(\frac{1}{2\eta}-\frac{L_f^\pi}{2}\right)\|\pi^{k+1}-\pi^k\|_F^2.
\end{align*}
To sum up, we get the desired result,
\begin{equation}
\label{eq:su_de}
\begin{aligned}
& \overline{F}(\pi^{k+1},\alpha^{k+1}) -\overline{F}(\pi^{k},\alpha^{k}) 
\\\le &  -\kappa_1\left(\|\pi^{k+1}-\pi^k\|_F^2 +\|\alpha^{k+1}-\alpha^k\|^2 \right), 
\end{aligned}
\end{equation}
where $\kappa_1 = \min\left(\frac{L_f^\alpha}{2}-\tfrac{1}{2\tau},\frac{L_f^\pi}{2}-\tfrac{1}{2\eta}\right) >0$ if $0<\tau<\tfrac{1}{L_f^\alpha}$ and $0<\epsilon<\tau<\tfrac{1}{L_f^\pi}$.
Summing up \eqref{eq:su_de} from $k=0$ to $+\infty$, we obtain 
\begin{equation}
\begin{aligned}
&\overline{F}(\pi^{\infty},\alpha^\infty)-\overline{F}(\pi^0,\alpha^0) \\ \leq& -\kappa_1 \sum_{k=0}^\infty\left(\|\pi^{k+1}-\pi^k\|_F^2+\|\alpha^{k+1}-\alpha^k\|^2\right). 
\end{aligned}
\end{equation}

As the potential function $\overline{F}(\cdot,\cdot)$ is bi-quadratic and thus coercive and $\{(\pi^{k},\alpha^k)\}_{k \ge 0}$ is a bounded sequence, it means the left-hand side is bounded, which implies
\begin{equation*}
\begin{aligned}
\sum_{k=0}^\infty &\left(\|\pi^{k+1}-\pi^k\|_F^2 +\|\alpha^{k+1}-\alpha^k\|^2\right) < +\infty,\\
&\alpha^{k+1}-\alpha^k \rightarrow 0, \pi^{k+1}-\pi^k \rightarrow 0.
\end{aligned}
\end{equation*}
 Let $(\pi^\infty, w^\infty)$ be a limit point of the sequence $\{(\pi^{k},\alpha^{k})\}_{k \ge 0 }$. Then, there exists a sequence $\{n_k\}_{k\ge0}$ such that $\{(\pi^{n_k},\alpha^{n_k})\}_{k \ge 0 }$ converges to $(\pi^\infty, \alpha^\infty)$. To proceed, we write down the optimality condition w.r.t \eqref{eq:alpha_update} and \eqref{eq:pi_update}, 
 \begin{equation}
 \label{eq:alpha_opt}
    0\in \nabla_\alpha F(\pi^{k},\alpha^k) + \frac{1}{\tau}(\alpha^{k+1}-\alpha^k) + \mathcal{N}_{\Theta}(\alpha^k),
 \end{equation}
 \begin{equation}
 \label{eq:pi_opt}
 0\in \nabla_\pi F(\pi^{k},\alpha^{k+1}) + \frac{1}{\eta}(\log(\pi^{k+1})-\log(\pi^k)) + \mathcal{N}_{C}(\pi^{k+1}), 
\end{equation}
 where $\mathcal{N}_C(x)$ is the normal cone of $C$ at the point $x$. 
 Replacing $k$ by $n_k$ in \eqref{eq:alpha_opt} and \eqref{eq:pi_opt}, 
 taking limits on both sides as $k\rightarrow\infty$ 
\begin{align*}
& 0\in \nabla_\alpha F(\pi^{\infty},\alpha^\infty)  + \mathcal{N}_{\Theta}(\alpha^\infty), \\
&  0\in \nabla_\pi F(\pi^{\infty},\alpha^{\infty})  + \mathcal{N}_{C}(\pi^{\infty}), 
\end{align*}
we obtain the desired result. 
\end{proof}

Intuitively, the result in Theorem \ref{thm} guarantees that the sequence $\{(\pi^k,\alpha^k)\}$ generated by SLOTAlign will converge to a critical point of the nonconvex optimization problem \eqref{eq:main_opt}.

\subsection{Complexity Analysis}\label{sec44}

Suppose $\mathcal G_s$ has $n_s$ nodes, $l_s$ edges, and $d_s$ node attributes while $\mathcal G_t$ has $n_t$ nodes, $l_t$ edges, and $d_t$ node attributes. We first consider the case of dense graphs. The complexity of candidate bases construction in \eqref{eq:proximity} is $O(n_s^2d_s+n_t^2d_t)$. The cost of calculating $F(\pi,\alpha)$ in \eqref{eq:main_opt2} is $O(n_s^2n_t+n_sn_t^2)$. The cost of $\alpha$-update in \eqref{eq:alpha_update} and $\pi$-update in \eqref{eq:pi_update} is $O(n_s^2n_t+n_sn_t^2)$. Therefore, the overall complexity is $O(n_s^2d_s+n_t^2d_t+n_s^2n_t+n_sn_t^2)$, which is the same order as other optimal transport based alignment methods \cite{xu2019gromov,chowdhury2021generalized,titouan2019optimal,li2022fast}.

If $\mathcal G_s$ and $\mathcal G_t$ are large-scale sparse graphs, i.e., $n_s^2\gg l_s$, $n_t^2\gg l_t$, $n_s\gg d_s$, and $n_t\gg d_t$, SLOTAlign can take advantage of the sparsity and low-rank properties of candidate structure bases  $\{\bm D^{(q)}_s\}_{q=1}^K$ and $\{\bm D^{(q)}_t\}_{q=1}^K$ in calculating \eqref{eq:main_opt2} and updating $\alpha$ and $\pi$. In this case, SLOTAlign can be optimized to quadratic time complexity $O(n_sn_t(d_s+d_t)+n_sl_t+n_tl_s)$.

In the experiments, SLOTAlign is able to align graphs with about 20,000 nodes efficiently (e.g., the DBP15K dataset). To further scale up SLOTAlign linearly, recent divide-and-conquer methods \cite{ge2021largeea, zeng2022entity} can be adopted. For example, LIME \cite{zeng2022entity} develops a bi-directional iterative graph partition strategy based on METIS \cite{karypis1998fast} to divide large-scale graph pairs into smaller subgraph pairs, and then applies alignment methods for each subgraph pair. It can preserve 80\% links when partitioning two graphs with millions of nodes into 75 subgraph pairs. Besides, LargeEA \cite{ge2021largeea} develops a mini-batch generation strategy to partition large graphs into smaller mini-batches for alignment. Therefore, SLOTAlign has great potential to be applied to graphs with millions of nodes. As the main target of this paper is to propose a more accurate and robust graph alignment method, we leave the scalability issue as our future work.

%% file: sec5_experiment.tex
\section{Experiments}\label{sec5}

\begin{figure*}[tb]
	\centering
	\includegraphics[width=\textwidth]{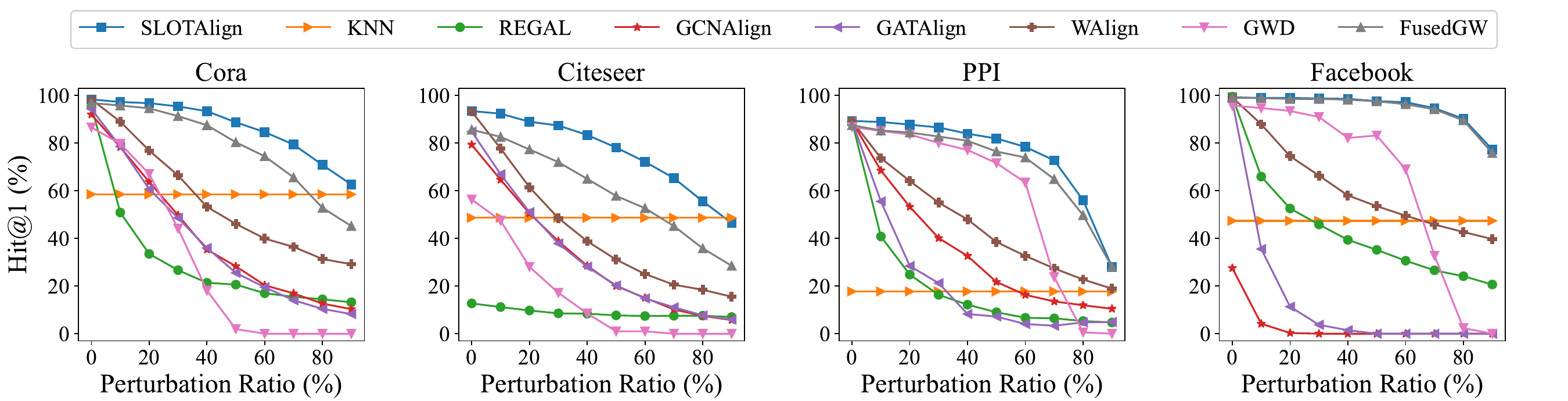} 
	\vspace{-4mm}
	\caption{Performance comparison of eight graph alignment methods under different levels of structure inconsistency.}
	\vspace{-3mm}
	\label{fig:exp1}
\end{figure*}

\begin{figure*}[tb]
	\centering
	\includegraphics[width=0.95\textwidth]{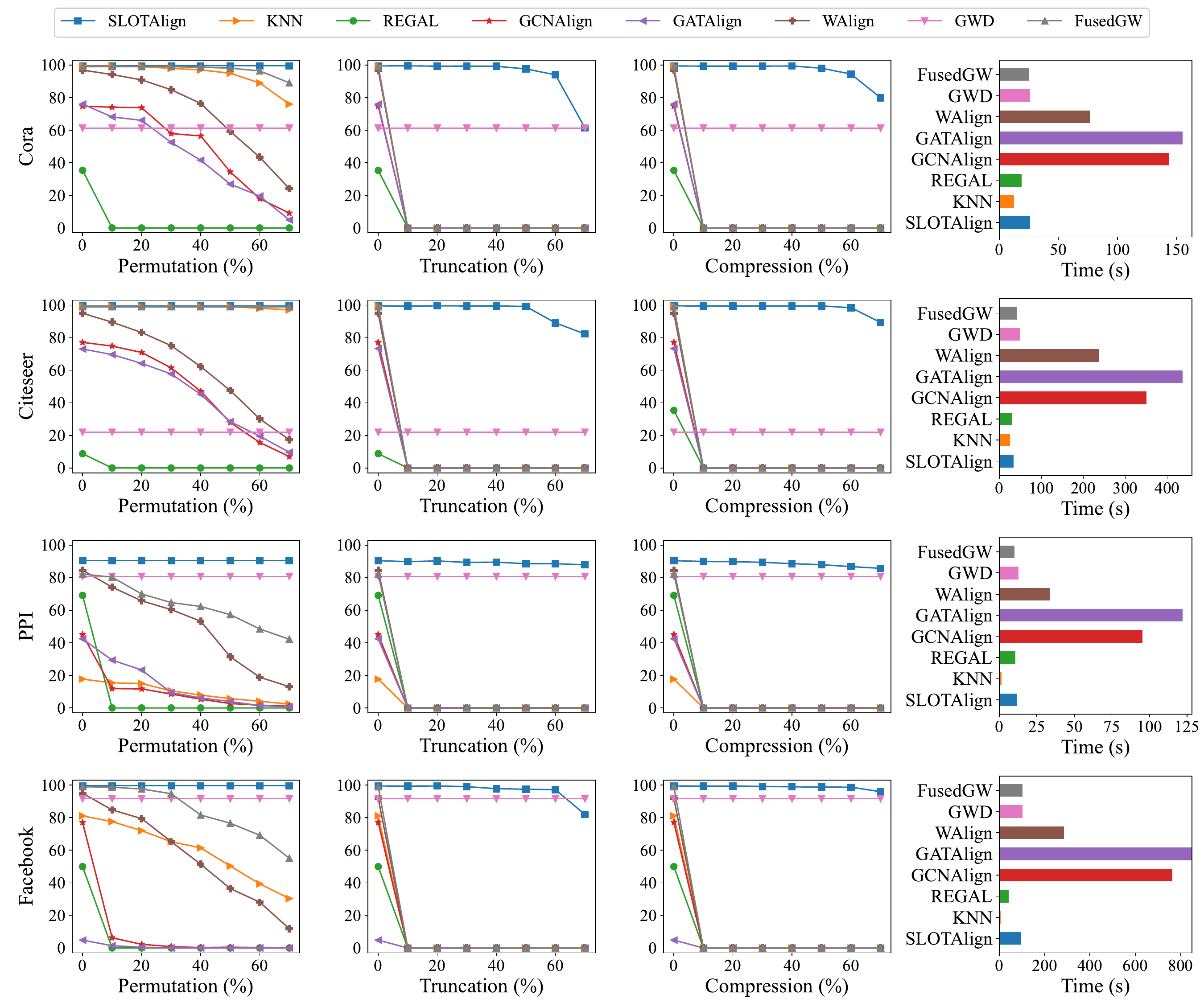} 
	\vspace{-3mm}
	\caption{Performance (Hit@1) and runtime comparisons of eight graph alignment methods under three types of feature inconsistency. }
	\vspace{-3mm}
	\label{fig:exp2}
\end{figure*}

In this section, we assess the effectiveness of the proposed SLOTAlign model. We aim to answer the following questions:
\begin{itemize}[leftmargin=*]
    \item (Q1) Is SLOTAlign more robust to feature and structure inconsistency compared with other alignment methods?
    \item (Q2) Does SLOTAlign outperform the state-of-the-art methods on noisy real-world graph alignment applications?
    \item (Q3) Does the joint structure learning and optimal transport framework in SLOTAlign really benefit the alignment performance?
    \item (Q4) How sensitive is SLOTAlign to different hyperparameters?
\end{itemize}
Our code and data are provided in the supplementary materials, and will be publicly available upon acceptance.

\subsection{Experimental Setup}\label{sec51}
We first introduce the experimental settings, including datasets, baseline methods, evaluation metrics, and implementation details.

\noindent\textbf{Datasets.} In Table \ref{tab:data}, we list the statistics of all datasets used in the experiments. Four semi-synthetic datasets with different degrees of structure and feature inconsistency are used to evaluate the robustness of SLOTAlign and baselines. Besides, three noisy real-world graph alignment datasets are used to assess the overall performance of all methods. The four semi-synthetic datasets are:
\begin{itemize}[leftmargin=*]
    \item \textbf{Cora} and \textbf{Citeseer} \cite{sen2008collective} are two citation networks in which nodes correspond to scientific publications and edges are citation links. Each publication node in the graph is described by a 0/1-valued word vector indicating the absence/presence of the corresponding word from the dictionary, i.e., the bag-of-words feature.
    \item \textbf{PPI} \cite{zitnik2017predicting} is a Protein-Protein Interaction network where nodes are proteins and edges indicate the interaction between proteins. Motif gene sets and immunological signatures are node features. 
    \item \textbf{Facebook}\cite{leskovec2012learning} is a social network where nodes represent Facebook accounts and edges reflect relations between accounts. Node features are extracted from user profiles.
\end{itemize}
For the above four datasets, we follow the experimental setup in \cite{heimann2018regal, chen2020cone}. We treat the original graph data as the source graph $\mathcal G_s=\{\mathcal U_s, \bm A_s, \bm X_s\}$, and perform node permutation to generate the corresponding target graph $\mathcal G_t=\{\mathcal V_t, \bm A_t, \bm X_t\}$ for alignment. Specifically, we have $\bm A_t=\bm P^T\bm A_s\bm P$ and $\bm X_t=\bm P^T\bm X_s$. Then we generate different levels of structure and feature inconsistency in $\mathcal G_s$. We introduce the detailed generation process in the next subsection. 

Subsequently, we evaluate all methods on the following noisy real-world graph alignment datasets without further modification: 
\begin{itemize}[leftmargin=*]
    \item \textbf{Douban Online-Offline.} In this scenario, we align two social network graphs, the online graph and the offline graph. In the online graph, nodes represent users, and edges represent the interaction between users (e.g., reply to a message) on the website. The offline graph is constructed according to the user's co-occurrence in social gatherings. The location of a user is used as node features in both graphs. The online graph is larger and contains all the users in the offline graph. In Douban, 1,118 users that appear in both graphs are used as the ground truth alignments.
    \item \textbf{ACM-DBLP.} In this scenario, we align two co-author networks ACM and DBLP, which are extracted from the publication information in four research areas. In both networks, nodes represent authors, and edges represent co-author relations. Node features indicate the number of papers that are published in different venues. There are 6,325 common authors across two networks used as the ground truth alignments.
    \item \textbf{DBP15K.} DBP15K\cite{sun2017cross} is a widely used KG alignment dataset. It consists of three cross-lingual entity alignment scenarios: DBP15K$_\textnormal{ZH\_EN}$ (Chinese to English), DBP15K$_\textnormal{JA\_EN}$ (Japanese to English), and DBP15K$_\textnormal{FR\_EN}$ (French to English). All three subsets are created from multi-lingual DBpedia, and each contains 15,000 pairs of aligned entities.  To preserve the feature inconsistency between multi-lingual knowledge graphs, we do not perform machine translation in data preprocessing. Instead, we follow \cite{liu2022selfkg} and use LaBSE \cite{LaBSE}, i.e., a multi-lingual BERT encoder, to extract 768-dimensional node features from each entity name.
\end{itemize}

\begin{table}[t]
\caption{Data Statistics. Attr. represents the number of node attributes.} 
\centering
\resizebox{\linewidth}{!}{
\begin{tabular}{l|cccc}
\toprule
Dataset & \# Nodes  & \# Edges & \# Attr. & Description \\\midrule
Cora \cite{sen2008collective}    & 2,708 & 5,028 & 1,433 & Citation Network  \\
Citeseer \cite{sen2008collective} & 3,327 & 4,732  & 3,703 & Citation Network  \\
PPI \cite{zitnik2017predicting}& 1,767 & 16,159 & 171 & Protein Interaction \\
Facebook \cite{leskovec2012learning}& 4,039 & 44,117 & 1,476 & Social Network \\\midrule
Douban \cite{zhang2016final} -Online & 3,906 & 16,328 & 538 & Social Network \\
\ \ \ \ \ \ \ \ \ \ \ \ \ \ \ \ -Offline & 1,118 & 3,022 & 538 & Social Network 
\\\midrule
ACM-DBLP\cite{zhang2018attributed} -ACM    & 9,872 & 39,561 & 17 & Co-Author Network \\
\ \ \ \ \ \ \ \ \ \ \ \ \ \ \ \ \ \ \ \ \ -DBLP   & 9,916 & 44,808 & 17 & Co-Author Network\\
\midrule
DBP15K$_\textnormal{ZH\_EN}$\cite{sun2017cross} -ZH    & 19,388   & 70,414   & 768  & Knowledge Graph \\
\ \ \ \ \ \ \ \ \ \ \ \ \ \ \ \ \ \ \ \ \ \ \ \ -EN    & 19,572   & 95,142   & 768  & Knowledge Graph \\
\midrule
DBP15K$_\textnormal{JA\_EN}$ -JA    & 19,814   & 77,214   & 768  & Knowledge Graph \\
\ \ \ \ \ \ \ \ \ \ \ \ \ \ \ \ \ \ -EN    & 19,780   & 93,484   & 768  & Knowledge Graph \\
\midrule
DBP15K$_\textnormal{FR\_EN}$ -FR    & 19,661   & 105,998  & 768  & Knowledge Graph \\
\ \ \ \ \ \ \ \ \ \ \ \ \ \ \ \ \ \ -EN    & 19,993   & 115,722  & 768  & Knowledge Graph \\
\bottomrule
\end{tabular}
}
\label{tab:data}
\end{table}

\noindent\textbf{Baselines.} We compare SLOTAlign with seven unsupervised graph alignment baselines, including KNN, four methods based on the ``embed-then-cross-compare'' paradigm (REGAL, WAlign, GCNAlign, and GATAlign) and two optimal transport based methods (GWD and FusedGW). We introduce these baselines as follows:
\begin{itemize}[leftmargin=*]
\item \textbf{KNN}. It is a simple baseline that directly matches nodes to top-k nearest neighbors in the feature space.
\item \textbf{REGAL} \cite{heimann2018regal}. It is a fast embedding-based graph alignment method that can be applied to graph with or without node features.
\item \textbf{WAlign} \cite{gao2021unsupervised}. It is a lightweight Graph Convolutional Network architecture with a Wasserstein distance discriminator to identify candidate node correspondences. These pseudo node correspondences are used to update network parameters and node embeddings.
\item \textbf{GCNAlign} \cite{wang2018cross}. It uses the Graph Convolutional Network \cite{Kipf} to calculate node embeddings, and synthesizes pseudo node correspondence pairs based on the cross-graph embedding similarity. The network is trained by the margin-based ranking loss, which makes corresponding nodes closer in the embedding space.
\item \textbf{GATAlign} \cite{velivckovic2017graph}. Its architecture is similar to GCNAlign mentioned above, but uses Graph Attention Network for node embedding learning.
\item \textbf{GWD} \cite{xu2019gromov}. Similar to SLOTAlign, it invokes GW distance for graph alignment. However, it merely considers the structural information and uses the graph adjacency matrix to represent the cost matrices. 
\item \textbf{FusedGW} \cite{titouan2019optimal}. It extends the GW distance to a new framework that takes into account both structure and feature information on graphs for the attributed graph alignment problem.
\end{itemize}

\noindent\textbf{Evaluation Metrics.} We use \textbf{Hit@k} to evaluate the performance of all graph alignment methods. It calculates the percentage of the nodes in $\mathcal V_t$ whose ground-truth alignment results in $\mathcal V_s$ is in the top-k candidates. We use all node correspondences across two graphs in evaluation. 

\noindent\textbf{Implementation Details.} For all mentioned baselines, we run the code provided by the authors and keep the default configuration. For SLOTAlign, we set the step size in the Sinkhorn algorithm $\eta$ as 0.01 by default in all datasets. The step size of structure learning $\tau$ is 0.1 in semi-synthetic datasets and 1 in real-world datasets. The number of candidate structure bases $K$ is 2 in semi-synthetic datasets and 4 in real-world datasets. Our model is implemented based on PyTorch and DGL \cite{wang2019dgl}.  All experiments are performed on a high-performance computing server running Ubuntu 20.04 with an AMD Ryzen9 5950X CPU and an NVIDIA GeForce RTX 3090 GPU.

\begin{table*}[t]
\centering
\caption{Experimental results and runtime of all compared methods on two real-world graph alignment scenarios.}\label{tab:exp}
\begin{tabular}{l|rrrrr|rrrrr}
\toprule
 & \multicolumn{5}{c|}{Douban Online-Offline} & \multicolumn{5}{c}{ACM-DBLP} \\
\midrule
Model      & {Hit@1} & {Hit@5} & {Hit@10} & {Hit@30} & {Time(s)} & {Hit@1} & {Hit@5} & {Hit@10} & {Hit@30} & {Time(s)} \\\midrule

KNN        & 3.31                      & 10.38                     & 16.64                      & 30.05                      & 0.9                      & 49.25                   & 59.46                   & 63.42                    & 69.61                    & 4.5                      \\\midrule
REGAL  &  *30.32 & *54.83  & - & - & - & 34.09 & 46.58 & 51.35 & 56.34 & 124.2 \\
GCNAlign & 20.93                     & 34.44                     & 39.62                      & 50.72                      & 248.3                    & 38.43                   & 68.46                   & 77.64                    & 86.89                    & 5,821.3                  \\
GATAlign & 23.70                     & 36.94                     & 44.01                      & 57.16                      & 264.5                    & 14.21                   & 34.07                   & 42.12                    & 49.00                    & 7,298.2                  \\
WAlign     & 35.69                     & 57.87                     & 69.05                      & 83.09                      & 129.6                    & 50.61                  & 72.87                   & 80.84                    & 89.47                    & 2,246.9                  \\\midrule
GWD        & 3.04                      & 7.96                      & 9.21                       & 11.90                      & 5.9                      & 56.24                   & 77.14                   & 82.20                    & 84.92                    & 269.8                    \\
FusedGW    & 29.61                     & 62.79                     & 66.46                      & 68.07                      & 99.6                     & 30.80                    & 38.39                   & 39.26                    & 39.6                     & 4,466.5                  \\\midrule
SLOTAlign  & \textbf{51.43}                     & \textbf{73.43}                     & \textbf{77.73}                     & \textbf{82.02}                      & 4.9                      & \textbf{66.04}                   & \textbf{84.06}                   & \textbf{87.95}                   & \textbf{90.32}                    & 234.5      \\
-w/o edge-view     & 2.42  & 10.02 & 16.37 & 32.11 & 4.1 & 30.42 & 48.16 & 53.26 & 58.59 & 224.6 \\
-w/o node-view     & 36.23 & 56.17 & 60.82 & 65.65 & 4.1 & 0.35  & 1.19  & 1.85  & 4.16  & 228.9 \\
-w/o subgraph-view & 22.09 & 35.15 & 40.43 & 45.97 & 3.6 & 65.75 & 83.84 & 87.65 & 90.01 & 162.0\\
-fixed $\beta_s$ and $\beta_k$  & 3.67  & 12.61 & 18.96 & 31.22 & 4.8 & 26.56 & 46.43 & 54.42 & 64.16 & 187.9 \\
-parameterized GNN & 40.34  & 56.62 & 60.55 & 68.43 & 7.2 & 64.27 & 81.83 & 85.23 & 87.62 & 540.6  \\
\bottomrule
\multicolumn{11}{l}{The results marked with * are obtained from \cite{gao2021unsupervised}. The rest of the results are reproduced by running the source code.}  
\vspace{-4mm}
\end{tabular}
\end{table*}

\subsection{(Q1) Alignment over Inconsistent Structures and Features}\label{sec52}

We first evaluate the robustness of the proposed SLOTAlign on four semi-synthetic datasets (Cora, Citeseer, PPI, and Facebook) against different degrees of structure inconsistency. To control the inconsistency level, we randomly perturb $\%p$ edges in $\mathcal G_t$ to other previous unconnected positions. In this setting, we only use the first 100 feature columns in Cora, Citeseer, and Facebook so that the models cannot only rely on node features for alignment. 

We show the results of all compared methods in Figure \ref{fig:exp1}. When the perturbation ratio is 0\%, SLOTAlign has comparable performance with other start-of-the-art graph alignment methods. When the structure perturbation ratio $p\%$ gradually increases from 0\% to 70\%, the performance of SLOTAlign degrades more slowly than other algorithms and consistently achieves the best in most cases. Note the maximum structure perturbation ratio in our setting is much larger compared with previous studies \cite{heimann2018regal, chen2020cone, gao2021unsupervised}. It validates that SLOTAlign is more resilient to structure inconsistency in graph alignment.

Subsequently, to evaluate the capability of each method in dealing with feature inconsistency, we consider three types of feature transformation on $\mathcal G_t$:
\begin{itemize}[leftmargin=*]
    \item \emph{Feature Permutation.} We randomly permute $p\%$ feature columns in $\mathcal G_t$, as defined in Definition \ref{def:PI}.
    \item \emph{Feature Truncation.} We randomly delete $p\%$ feature columns in $\mathcal G_t$.
    \item \emph{Feature Compression:} We use the Principal Component Analysis (PCA) to compress features in $\mathcal G_t$ into a low-dimensional representation with compression ratio $p\%$.
\end{itemize}
These transformations can be regarded as three basic feature inconsistency simulators in real-world applications. For example, the Cora graph uses bag-of-words as node features to indicate the presence of a word in the document. First, consider that bag-of-words features in two graphs are built on the same vocabulary set but in different orders. This unaligned scenario can be fully characterized by the feature perturbation simulator. Second, feature truncation simulates a scenario where the vocabulary used in $\mathcal G_t$ is a subset of that in $\mathcal G_s$. Third, feature compression in this example can be interpreted as the alignment between sparse bag-of-words features in $\mathcal G_s$ and dense low-dimensional features in $\mathcal G_t$. 

For each transformation type, we gradually increase the inconsistency ratio $p\%$ from 0\% to 70\%. We perturb 25\% edges simultaneously to ensure that the models cannot align two graphs purely based on the structural information. We report the experimental results of SLOTAlign and other baselines in Figure \ref{fig:exp2} and analyze the model performance from the following perspectives.

\noindent\textbf{(1) Effect of Feature Permutation.} In the first column of Figure \ref{fig:exp2}, we observe that the feature permutation has no influence on our proposed SLOTAlign, consistent with our proof in Proposition \ref{prop:1}. In comparison, other methods using node features are significantly affected. For example, the Hit@1 of all baselines except GWD decreases to lower than 50\% on the PPI dataset when the permutation ratio increases to 70\%. Although the structure-based approach GWD is not affected by any type and degree of feature inconsistency, its performance is also significantly lower than SLOTAlign as the feature information is not utilized.

\noindent\textbf{(2) Effect of Feature Truncation and Compression.} In the second and third columns of Figure \ref{fig:exp2}, the performance of SLOTAlign keeps stable if the truncation or compression ratio is less than 50\%. Even though the ratio is greater than 50\%, SLOTAlign still outperforms the structure-based approach GWD, which verifies the robustness of SLOTAlign against different types and levels of feature inconsistency. On the contrary, other baselines using graph features fail to align two graphs with any level of feature truncation or compression. As we analyzed in Section \ref{sec3}, these unsupervised methods are unable to perform cross-graph node embedding comparison if the embedding spaces of two graphs are not aligned.

\noindent\textbf{(3) Efficiency Comparison.} In the last column of Figure \ref{fig:exp2}, we compare the computational efficiency of each method. REGAL has the shortest running time, but also the worst performance. SLOTAlign, GWD, and fusedGW have comparable running time as all of them are GW-based methods. Compared with graph neural network based methods WAlign, GCNAlign, and GCNAlign, SLOTAlign is more efficient on all datasets.

\subsection{(Q2) Alignment on Real-world Graphs}\label{sec54}

Next, we evaluate all methods on two noisy real-world graph alignment datasets, namely Douban Online-Offline and ACM-DBLP. The experimental results and runtime are reported in Table \ref{tab:exp}. Our proposed SLOTAlign achieves the best performance in terms of the alignment accuracy on two datasets. Specifically, SLOTAlign has 15.7\% and 15.4\% absolutely improvement in Hit@1 on Douban and ACM-DBLP, respectively, compared with the state-of-the-art unsupervised alignment method WAlign. The performance improvement is also very significant compared with GW-based methods (GWD, FusedGW), proving the superiority of SLOTAlign. Besides, SLOTAlign is time-efficient and only slower than two classic methods, KNN and REGAL.

\begin{table}[t!]
\centering
\caption{Evaluation Results of all compared knowledge graph alignment methods on DBP15K.}
\label{tab:kg}
\resizebox{\linewidth}{!}{
\begin{tabular}{l|rr|rr|rr}\toprule
          & \multicolumn{2}{|r|}{DBP15K$_\textnormal{ZH\_EN}$}& \multicolumn{2}{r|}{DBP15K$_\textnormal{JA\_EN}$} & \multicolumn{2}{r}{DBP15K$_\textnormal{FR\_EN}$} \\
Method    & Hit@1           & Hit@10          & Hit@1           & Hit@10          & Hit@1           & Hit@10          \\\midrule
GCNAlign  & 43.4            & 76.2            & 42.7            & 76.2            & 41.1            & 77.2            \\
LIME      & 87.4            & -               & 90.9            & -               & 97.8            & -               \\\midrule
MultiKE   & 50.9            & 57.6            & 39.3            & 48.9            & 63.9            & 71.2            \\
EVA       & 75.2            & 89.5            & 73.7            & 89.0            & 73.1            & 90.9            \\
SelfKG    & 74.5            & 86.6            & 81.6            & 91.3            & 95.7            & 99.2            \\
SLOTAlign & \textbf{89.0}   & \textbf{94.4}            & \textbf{93.0}   & \textbf{96.5}            & \textbf{99.2}   & \textbf{99.8}            \\
\bottomrule
\end{tabular}
}
\vspace{-4mm}
\end{table}

We also evaluate SLOTAlign on the DBP15K dataset for KG alignment. To reduce the difficulty of large-scale optimization, we initialize $\pi^1$ with the node-wise feature similarity matrix instead of the uniform distribution. We compare SLOTAlign with two supervised methods ---  GCNAlign \cite{wang2018cross} and LIME \cite{zeng2022entity}, and three unsupervised methods ---  MultiKE \cite{MultiKE}, EVA \cite{EVA}, and SelfKG \cite{liu2022selfkg}.  The experimental results are summarized in Table \ref{tab:kg}. Our proposed SLOTAlign achieves best performance on all the metrics. It corroborates our theoretical insights that SLOTAlign can better deal with the feature inconsistency issue in multi-lingual KG alignment. Since SLOTAlign focuses on the general attributed graph alignment problem, we do not consider the additional information in KG (e.g., relation type, entity description, and machine translation), even though they may further boost the performance \cite{li2022uncertainty, ding2022conflict}.

\begin{figure}[t]
	\centering
	\includegraphics[width=\linewidth]{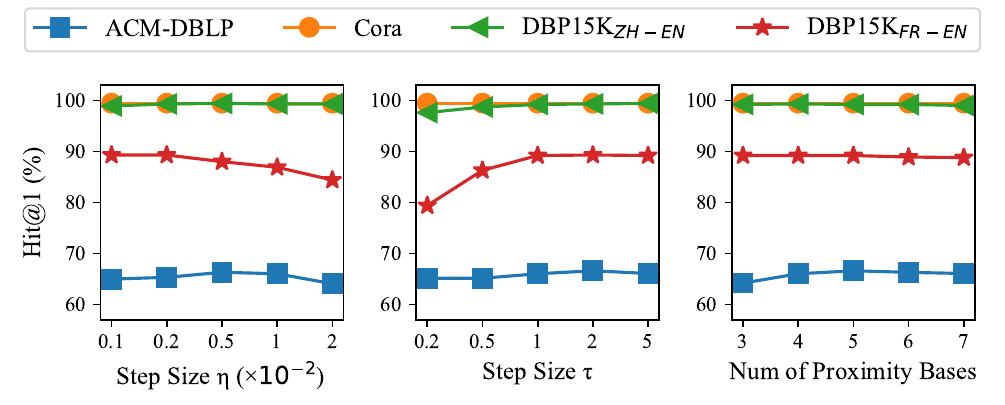} 
	\vspace{-7mm}
	\caption{Sensitivity analysis of SLOTAlign on four datasets.}
	\vspace{-3mm}
	\label{fig:exp3}
\end{figure}

\subsection{(Q3, Q4) Analysis of SLOTAlign}\label{sec56}
\noindent\textbf{Ablation Study.} To validate the effectiveness of each component in SLOTAlign, we compare it with several ablations. After removing each of the three views in the multi-view modeling, we obtain SLOTAlign without (w/o) edge-view, node-view, and subgraph-view. Besides, we conduct the ablation that does not perform structure learning, i.e., keep the weights of candidate structure bases $\beta_s$ and $\beta_t$ fixed in the optimization process. As shown in the bottom block of Table \ref{tab:exp}, SLOTAlign consistently achieves the best performance compared with these variants, which validates the effectiveness of the proposed multi-view structure modeling and joint learning framework. In the last row of Table \ref{tab:exp}, we compare the parameter-free GNN in subgraph-view with the original parameterized GNN \cite{wu2019simplifying} with linear layers and the ReLU activation function. We use the same loss in Equation \eqref{eq:main_opt2} for GNN training. Although the parameterized GNN may have better expressive power in embedding-based alignment methods \cite{sun2020benchmarking, liu2022selfkg}, we find that it has inferior performance compared to the parameter-free GNN in SLOTAlign.

\noindent\textbf{Sensitivity Analysis of SLOTAlign.} We investigate the sensitivity of SLOTAlign to hyperparameters in Algorithm \ref{alg}. On four datasets including ACM-DBLP, Cora, DBP15K$_\textnormal{ZH-EN}$, and DBP15K$_\textnormal{FR-EN}$, we analyze the performance of SLOTAlign with different step sizes of structure learning $\tau \in \{0.2, 0.5, 1, 2, 5\}$, different step sizes in the Sinkhorn algorithm $\eta \in \{0.001, 0.002, 0.005, 0.01, 0.02\}$, and different number of structure bases $K=\{3,4,5,6,7\}$. As shown in Figure \ref{fig:exp3}, the performance of SLOTAlign is typically robust to all these hyperparameters. Without hyperparameter tuning, the default configuration (i.e., $\eta=0.01, \tau=1, K=4$) is sufficient to get competitive results on these datasets.

%% file: sec6_relatedwork.tex
\section{Related Work}\label{sec6}

\textbf{Supervised graph alignment} algorithms use a set of known seed node pairs between graphs to infer node correspondences. COSNET \cite{zhang2015cosnet} uses an energy-based model to describe both global and local consistency. In addition, many embedding-based methods have been proposed to learn node embeddings and then predict corresponding node pairs \cite{liu2016aligning, man2016predict, zhou2018deeplink}. For example, CrossMNA \cite{chu2019cross} considers inter-vector and intra-vector to combine graph and node embeddings, respectively. \cite{fey2019deep} propose a two-stage neural architecture to learn node embeddings and match nodes between graphs. ATTENT \cite{zhou2021attent} utilizes active learning to improve graph alignment with limited seed pairs. BRIGHT \cite{yan2021bright} uses anchor links between seed pairs as landmarks to construct a certain unified space for matching by random walk. NEXTALIGN \cite{zhang2021balancing} reveals the close connections between graph convolutional networks and consistency-based alignment methods, and strikes a balance between the alignment consistency and disparity. Most existing knowledge graph (KG) alignment methods \cite{trisedya2019entity, LaBSE, MultiKE, EVA, zeng2022entity, ge2021largeea, liu2022selfkg, mao2020relational, sun2017cross, wang2018cross, sun2020benchmarking} also follow the ``embed-then-cross-compare'' paradigm. For example, AttrE \cite{trisedya2019entity} proposes the attribute character embeddings which shifts the entity embeddings from two KGs into the same space for similarity calculation. LIME \cite{zeng2022entity} learns the unified entity representations using a reciprocal alignment inference strategy to model the bi-directional entity interactions across graphs

\textbf{Unsupervised graph alignment} methods predict node correspondences across graphs without the requirement of any labeled node pairs. It has attracted increasing attention as node pairs are usually unavailable in real-world scenarios \cite{gao2021unsupervised}. To solve this problem, FINAL \cite{zhang2016final} proposes the consistency principle, i.e., if two pairs of nodes are similar, their alignments should be consistent. HashAlign \cite{heimann2018hashalign} leverages locality sensitive hashing to help node matching based on the node similarity obtained from graph structure and node features. REGAL \cite{heimann2018regal} jointly obtains node representations from multiple graphs by factorizing matrix and matches the most similar node embedding across graphs. CONE-Align \cite{chen2020cone} proposes a proximity-preserving node embedding method to make different graphs comparable. GRASP \cite{hermanns2021grasp} considers the graph alignment problem as a mapping between functions on graphs, and the node embeddings are interpreted as the linear combinations of eigenvectors. Kyster et al. \cite{kyster2021boosting} develop an enhanced algorithm variant based on REGAL, CONE-Align and GRASP. Karakasis et al. \cite{karakasis2021joint} propose to learn the node embeddings and matches the nodes of two graphs simultaneously. Considering the big success in various graph mining tasks, graph neural networks are integrated in graph alignment algorithms to acquire better node representations \cite{gao2021unsupervised, liang2021unsupervised}, and adversarial learning strategies are utilized to further improve model performance \cite{chen2019unsupervised, derr2021deep}. Besides, many unsupervised methods have been proposed for KG alignment \cite{liu2022selfkg, mao-etal-2021-alignment}. For example, SelfKG \cite{liu2022selfkg} uses the graph attention network to aggregate entity embeddings of one-hot neighbors, and proposes a relative similarity metric between the entities of two KGs for self-supervised contrastive learning.